%% file: main.tex
\documentclass[10pt,twocolumn,twoside]{IEEEtran}
\input{main-packages.tex}

\begin{document}
\title{A Distributed Scheme for Stability Assessment in Large-Scale Structure-Preserving Models via Singular Perturbation}

\author{Amin~Gholami,~\IEEEmembership{Student Member,~IEEE}~and~Xu~Andy~Sun,~\IEEEmembership{Senior Member,~IEEE}
	\thanks{The authors are with the H. Milton Stewart School of Industrial and Systems Engineering, Georgia Institute of Technology, Atlanta, GA 30332 USA (e-mail: \texttt{a.gholami{@}gatech.edu}; \texttt{andy.sun{@}isye.gatech.edu}).}}
\maketitle

\begin{abstract}
Assessing small-signal stability of power systems composed of thousands of interacting generators is a computationally challenging task. To reduce the computational burden, this paper introduces a novel condition to assess and certify small-signal stability. Using this certificate, we can see the impact of network topology and system parameters (generators’ damping and inertia) on the eigenvalues of the system. The proposed certificate is derived from rigorous analysis of the classical structure-preserving swing equation model and has a physically insightful interpretation related to the generators’ parameters and reactive power. To develop the certificate, we use singular perturbation techniques, and in the process, we establish the relationship between the structure-preserving model and its singular perturbation counterpart. As the proposed method is fully distributed and uses only local measurements, its computational cost does not increase with the size of the system. The effectiveness of the scheme is numerically illustrated on the WSCC system.
%

%
%
\end{abstract}

\IEEEpeerreviewmaketitle

\section{Introduction}
\subsection{Motivation}
The rapid growth of renewable energy sources, open access transmission, intensifying competition in electricity markets, and aging transmission infrastructure are reshaping the operation of power systems in new ways that raise unprecedented challenges to the stability of the power grid. Mitigating power system instability would be a real challenge for power system operators. The advent of wide area measurement system could pave the way for improving the situational awareness of system operators and set the stage for new ways of stability assessment in power systems. Nonetheless, there is an urgent need for developing novel methods that combine the classical model-based approaches with the new measurement-based ones in order to achieve faster stability monitoring and assessment. This paper is motivated by this urgent need and aims to develop a fully distributed control scheme for the small-signal stability of the structure-preserving swing equation model of power systems.  

\subsection{Related works}
Broadly speaking, the vast literature on power system stability can be classified based on two modeling assumptions. The first and the more classical one assumes the stability model under study is fully known, whereas the second stream of research is model-free and adopts synchronized wide-area measurements in order to monitor and address the stability problem \cite{2017-Mani-stability-measurement}, \cite{2011-follum-evaluation}. In this measurement-based approach, the underlying model of the system is not necessarily known. Our work in the present paper is an attempt to combine the measurement-based and model-based approaches with the aim of achieving faster stability assessment.

 
Considering the model-based approach, the classical model for rotor angle stability analysis is the swing equation \cite{2020-fast-certificate}. This model is based on representing loads as constant impedances, and then incorporating load impedances into the nodal admittance matrix for a reduced network with only generator buses. Much effort has been devoted to understanding the stability properties of this network-reduced model, e.g. studying its small-signal stability \cite{2020-fast-certificate}, hyperbolicity and bifurcation \cite{2020-role-of-damping}, phase portrait \cite{1988-Zaborszky-phase-portrait}, constructing energy functions and Lyapunov functions \cite{1983-Vittal-power}, and using direct methods \cite{2011-Chiang-book-direct-methods}.
 
Among the various simplifying assumptions applied to the swing equation \cite{2008-anderson-stability}, ignoring the transfer conductance of the transmission lines and load dynamics are the most unrealistic ones. In 1981, Bergen and Hill introduced their well-known structure-preserving model \cite{1981-Bergen-Hill-Structure-Preserving} for the swing equation. The main assumption of this model is to use a frequency-dependent model for loads. In the present paper, we base our stability analyses upon this structure-preserving model. The structure-preserving model leads to more realistic analyses, and since its introduction, many researchers have based their investigations upon it. For instance, in \cite{2011-Dorfler-topological}, Dorfler \textit{et al.} show that locally near the synchronization manifold, the phase and frequency dynamics of the Bergen and Hill network-preserving model are topologically conjugate to the phase dynamics of a nonuniform Kuramoto model together with decoupled and stable frequency dynamics. In \cite{2016-Turitsyn-framework-structure-preserving}, the transient stability problem in a structure-preserving model is addressed using the quadratic Lyapunov functions approach.

The use of a network-preserving model enables us to study the impact of network topology and system parameters on the system stability. In this regard, a related study is \cite{2019-kasra-input-swing}, where the classical network-reduced swing equation model is used to examine how the network topology (i.e., the reduced fictitious network) will affect the system transfer function.


In this paper, we tackle the small-signal stability problem. Recall that \emph{small-signal stability} concerns with the ability of a power system to maintain generator phase synchronism under small disturbances \cite{2004-kundur-definition}. There is a large body of work on the model-based small-signal stability assessment \cite{2019-wilches-damping,2019-wilches-small-signal-hicss,2020-chow-small-signal-book,2017-milano-primal-small-signal,1980-Skar-stability-thesis}. For instance, \cite{2019-wilches-damping} proposes a method to damp inter-area oscillations using system loads, and \cite{2019-wilches-small-signal-hicss} examines the role of wind turbine integration in these inter-area oscillations. In \cite{2017-milano-primal-small-signal}, small-signal stability of power systems is investigated based on matrix pencils and the generalized eigenvalue problem. The paper compares different formulations and the state-of-the-art solvers. Finally, in \cite{1980-Skar-stability-thesis} it is shown that unstable equilibrium solutions for swing equations may exist even though the rotor angles are less than $90^\circ$ out of phase.



\subsection{Main results and paper outline}

In this paper, we combine the measurement-based and model-based approaches to develop a condition that certifies the small-signal stability of a structure-preserving swing equation model. The proposed certificate is a practical alternative to the eigenvalue computation-based methods, which can be quite computationally cumbersome for large-scale systems. We also introduce a control scheme for improving the system small-signal stability.  

The proposed control and assessment schemes can be implemented in a completely distributed fashion and do not require any information exchange between the neighboring generators and areas. This property makes them particularly suitable for fast assessment in large-scale power systems and when proprietary information from neighboring areas or power plants cannot be shared. 

In the process, we investigate the impact of network topology and system parameters (generator's inertia and damping) on the stability of the system. We introduce an stability index which provides a
quantitative measure of the degree of stability.

We make use of singular perturbation techniques to establish the relationship between a structure-preserving model and its singular perturbation counterpart. Specifically, we show (under specific conditions) the stability properties of the structure-preserving model are the same as those of its singular perturbation counterpart. Therefore, the singular perturbation counterpart can be used for small-signal analysis instead of the the structure-preserving model, and this will facilitate our analysis.

The rest of our paper is organized as follows. Section \ref{sec: Power system model} introduces the structure-preserving model as well as its singular perturbation counterpart. The main results of our paper are presented in Section \ref{sec: Main results}. Section \ref{sec: Numerical experiments} further exhibits the validity and conservativeness of the proposed stability certificate. Finally, the paper concludes with Section \ref{sec: Conclusions}.

\section{Power system model}
\label{sec: Power system model}
The classical swing equation model is based on a set of simplifying assumptions (see \cite{2008-anderson-stability} for the details of the assumptions) out of which ignoring the transfer conductances is the most unrealistic one. This issue stems from the fact that the loads are considered constant impedances and reflected into the nodal admittance matrix which will be further reduced to a smaller matrix representing a reduced network of generator buses. Ignoring the real part of this reduced admittance matrix seems, therefore, unreasonable since this real part is not only representing the resistive part of the transmission lines, but also the active power consumption of the system. Aside from this, the procedure of network reduction for eliminating the load buses will close our eyes on the relations between the structure of the underlying network and the stability properties of the system. With these in mind, the small-signal stability analysis in this paper is based on the standard structure-preserving model \cite{1981-Bergen-Hill-Structure-Preserving}. This model incorporates the nonlinear swing equation dynamics of generators as well as the frequency-dependent dynamics of loads. The model also preserves the original network topology (rather than undergoing the usual Kron reduction). We will exploit this preservation of topology later to analyze the effect of network topology on the stability of the system.

\subsection{Structure-preserving model}
Since generators are connected to the network through transient reactances, it is convenient to introduce fictitious buses representing the internal generator voltages, and further consider the transient reactances to be a transmission line. In the sequel, we assume that this transformation has been done, and therefore, the buses of the network can be categorize into generator buses (internal generator buses) and load buses. Moreover, with this transformation, no load is connected to generator buses and vice versa.

Consider an $n$-bus power system for some $n\in\mathbb{N}$ with the set of transmission lines $\mathcal{E}$. Let $ \mathcal{G}=\{1,\cdots, n_0\}$ be the set of generator buses, and $ \mathcal{L}=\{n_0+1,\cdots,n\}$ be the set of load buses. Based on the classical small-signal stability assumptions \cite{1981-Bergen-Hill-Structure-Preserving}, the structure-preserving model of this power system is
\begin{subequations} \label{eq: Swing Equation Net Preserving}
	\begin{align}
	  {m_i} \ddot{\delta}_i +  {d_i} {\dot{\delta}}_i &= P_{m_i}  - P_{e_i}  && \forall i \in \mathcal{G}, \label{eq: swing eq second order 1}     \\
	 d_i \dot{\delta}_i &= - P_{d_i} - P_{e_i}  && \forall i \in \mathcal{L}, \label{eq: swing eq second order 2}
	\end{align}
\end{subequations}
where \eqref{eq: swing eq second order 1} and \eqref{eq: swing eq second order 2} characterize the the dynamics of generator buses and load buses, respectively. In these equations, $\delta_i$ is the bus voltage angle in radians. For each generator bus $i\in\mathcal{G}$, $P_{m_i}$ is the mechanical power in per unit. Moreover, $m_i = \frac{M_i}{\omega_s}$ and $d_i = \frac{D_i}{\omega_s}$, where $M_i>0$ is the inertia constant in seconds, $D_i>0$ is the unitless damping coefficient, and $\omega_{s}$ is the synchronous angular velocity in electrical radians per seconds.

For each load bus $i\in\mathcal{L}$, ${d}_i>0$ is the frequency-dependence coefficient and $P_{d_i}$ is the load value in per unit at the current operating point.

%
In general, the real power drawn by load $i\in\mathcal{L}$ is a nonlinear function of voltage and frequency. Under small-signal stability assumptions, voltages are constant, and for small frequency variations around an operating point $P_{d_i}$, it is reasonable to consider the frequency-dependent load model
\begin{align} \label{eq: load model}
    - P_{e_i} = P_{d_i} + {d}_i \dot{\delta}_i && \forall i \in \mathcal{L}.
\end{align}
This load model describes the dynamics at load buses in \eqref{eq: swing eq second order 2}. Note that as ${d}_i\to 0$ in \eqref{eq: load model}, we approach a constant-power load model. In \eqref{eq: Swing Equation Net Preserving} and \eqref{eq: load model}, $P_{e_i}$ is the active electrical power injected from bus $i$ into the network, and is given by
\begin{align} \label{eq: flow function}
P_{e_i}  = \sum \limits_{j = 1}^n { V_i  V_j Y_{ij} \cos \left( \theta _{ij} - \delta _i + \delta _j \right)},
\end{align}
where $V_i $ represents the voltage magnitude of the $i$th bus which is assumed to be constant, and $Y_{ij}\angle \theta _{ij} $ is the $(i,j)$th entry of the nodal admittance matrix.
\subsection{Equilibrium points}
\label{subsec: Steady state}
The \emph{state} of system \eqref{eq: Swing Equation Net Preserving} is characterized by the vector $x=[\delta_1,\cdots,\delta_n,\dot{\delta}_1,\cdots, \dot{\delta}_{n_0}]^\top$. An \emph{equilibrium point} of system \eqref{eq: Swing Equation Net Preserving} is a state $x^*$ such that if the system reaches $x^*$, it will stay there permanently. Particularly, in \eqref{eq: Swing Equation Net Preserving}, an equilibrium point $x^*$ is of the form $x^*=[\delta_1^*,\cdots,\delta_n^*,0,\cdots, 0]^\top$. Indeed, the generator frequency deviations are zero, i.e., $ \dot{\delta}_i^*=0, \forall i \in \mathcal{G}$, and the set of bus angles $\delta_i^*, \forall i \in \mathcal{G}\bigcup \mathcal{L}$ is a solution to the following system of active power flow equations  
\begin{align*}
 P_{m_i}  =& \sum \limits_{j = 1}^n { V_i  V_j Y_{ij} \cos \left( \theta _{ij} - \delta_i^* + \delta_j^* \right)} && \forall i \in \mathcal{G}, \\
 -P_{d_i} =& \sum \limits_{j = 1}^n { V_i  V_j Y_{ij} \cos \left( \theta _{ij} - \delta_i^* + \delta_j^* \right)} && \forall i \in \mathcal{L}.
\end{align*}
Note also that solution of the above active power flow equations is not unique since any shift $c$ in the bus angles, i.e., $\delta_i^*+c, \forall i \in \mathcal{G}\bigcup \mathcal{L}$ is also a solution. However, this translational invariance can be dealt with by defining a reference bus and referring all other bus angles to it.
\begin{assumption} \label{assump}
An equilibrium point of system \eqref{eq: Swing Equation Net Preserving} satisfies the condition $0< (\theta _{ij} - \delta _i^* + \delta _j^*) < \pi$ for all transmission lines $(i,j)\in\mathcal{E}$.
\end{assumption}
Recall that this is a reasonable assumption since the entries of the admittance matrix, i.e., $Y_{ij} \angle \theta_{ij}$ satisfy $\frac{\pi}{2}\le \theta_{ij} < \pi, \forall (i,j)\in\mathcal{E}$. In lossless networks, we have $\theta_{ij}=\frac{\pi}{2}$, and thus Assumption \ref{assump} translates to $|\delta _i^* - \delta _j^*|<\frac{\pi}{2}$. More generally, the X/R ratio, i.e., the ratio of the line reactance to the line resistance is significantly above unity in lossy transmission networks. Therefore, $\theta_{ij}$ is close to $\frac{\pi}{2}$, and Assumption \ref{assump} translates to $|\delta _i^* - \delta _j^*|< \gamma <\frac{\pi}{2}$, for some number $\gamma$ close to $\frac{\pi}{2}$.

\subsection{Singular perturbation model}
To facilitate the analysis, we study the \emph{singular perturbation model} of dynamical system \eqref{eq: Swing Equation Net Preserving}: 
\begin{subequations} \label{eq: Swing Equation Net Preserving singular perturbation}
	\begin{align}
	 m_i \ddot{\delta}_i +  d_i {\dot{\delta}}_i &= P_{m_i}  - P_{e_i}  && \forall i \in \mathcal{G},  \\
	 {\varepsilon} \ddot{\delta}_i + {{d}}_i \dot{\delta}_i &= - P_{d_i} - P_{e_i}  && \forall i \in \mathcal{L}, \label{eq: Swing Equation Net Preserving singular perturbation loads}
	\end{align}
\end{subequations}
where the variables $\ddot{\delta}_i,\forall i \in \mathcal{L}$ are multiplied by a \emph{small positive parameter} $\varepsilon$. Note that by setting $ \varepsilon=0$ we will return to the original unperturbed model \eqref{eq: Swing Equation Net Preserving}. The main motivation for working with the singular perturbation model \eqref{eq: Swing Equation Net Preserving singular perturbation} in this paper is that it will pave the way for developing a stability certificate for the equilibrium points. Naturally, it is important to find the relationship between this auxiliary model \eqref{eq: Swing Equation Net Preserving singular perturbation} and the original structure-preserving model \eqref{eq: Swing Equation Net Preserving}. If the two models have the same stability properties, then it is reasonable to work with the model that is easier to analyze. We will see if this is the case in the next section.

Note that, the form of the equilibrium points of the singular perturbation system \eqref{eq: Swing Equation Net Preserving singular perturbation} is similar to those of system \eqref{eq: Swing Equation Net Preserving}, discussed in Section \ref{subsec: Steady state}. Likewise, Assumption \ref{assump} can be applied to system \eqref{eq: Swing Equation Net Preserving singular perturbation}.

\section{Main results: A stability certificate}
\label{sec: Main results}
Three questions naturally arise regarding the equilibrium points of systems \eqref{eq: Swing Equation Net Preserving} and \eqref{eq: Swing Equation Net Preserving singular perturbation}:
\begin{enumerate}[Q1]
    \item Which equilibrium points are stable? \label{Q1}
    \item What is the relationship between the stability of an equilibrium  point and the parameters (e.g., damping, inertia, network topology, etc.) of the system? \label{Q2}
    \item What is the relationship between the stability of system \eqref{eq: Swing Equation Net Preserving} and its singular perturbation counterpart \eqref{eq: Swing Equation Net Preserving singular perturbation}? \label{Q3}
\end{enumerate}

Obviously, \textcolor{mygreen}{Q}\ref{Q1} can be addressed by finding the eigenvalues of the Jacobian matrix associated with the first-order representation of the system (see \eqref{eq: SE NP reduced} and \eqref{eq: Swing Equation Net Preserving first order singular perturbation} in Appendix \ref{appen: proof of theorems} for more details). Another possibility is to numerically construct a Lyapunov function for this system using semidefinite programming techniques. Clearly, both of these ways are computationally expensive and not applicable to realistic large-scale systems. Theorem \ref{thrm: main result} provides a computationally tractable condition to certify the stability of an equilibrium point, therefore, provides an answer to \textcolor{mygreen}{Q}\ref{Q1}. Incidentally, this theorem also tackles \textcolor{mygreen}{Q}\ref{Q2}.
\begin{theorem} \label{thrm: main result}
Consider the singular perturbation model \eqref{eq: Swing Equation Net Preserving singular perturbation} with an equilibrium point $x^*$ that satisfies Assumption \ref{assump}. If the condition
\begin{align} \label{eq: condition for stability}
            -Q_i - V_i^2B_{ii} \le \frac{d_i^2}{2m_i}     && \forall i \in \mathcal{G},
\end{align}
is satisfied, then the equilibrium point is locally asymptotically stable. In \eqref{eq: condition for stability}, $Q_i$ denotes the reactive power injected from bus $i$ into the network, given by
\begin{align*}
    Q_i  = -\sum \limits_{j = 1}^n { V_i  V_j Y_{ij} \sin \left( \theta_{ij} - \delta _i^* + \delta_j^* \right)}.
\end{align*}
Furthermore, $B_{ii}$ is the imaginary part of the $i$th diagonal element of the bus admittance matrix.
\end{theorem}
 Next, Theorem \ref{thrm: relation} answers \textcolor{mygreen}{Q}\ref{Q3}. This theorem justifies the use of singular perturbation for stability analysis. Recall that an equilibrium point is hyperbolic, if the Jacobian of the corresponding first-order system has no eigenvalues on the imaginary axis.
\begin{theorem} \label{thrm: relation}
Consider the structure-preserving model \eqref{eq: Swing Equation Net Preserving} and its singular perturbation counterpart \eqref{eq: Swing Equation Net Preserving singular perturbation}. The following statements hold:
\begin{enumerate}[(i)]
    \item If $x^*$ is an exponentially stable equilibrium point of the unperturbed model \eqref{eq: Swing Equation Net Preserving}, then the corresponding equilibrium point of the singular perturbation model \eqref{eq: Swing Equation Net Preserving singular perturbation} is also exponentially stable, for sufficiently small $\varepsilon$. \label{i of thrm: relation}
    \item Suppose for every sufficiently small $\varepsilon$, $y^*$ is an asymptotically stable equilibrium point of the singular perturbation model \eqref{eq: Swing Equation Net Preserving singular perturbation}. If $x^*$ is a corresponding hyperbolic equilibrium point of the unperturbed model \eqref{eq: Swing Equation Net Preserving}, then $x^*$ is also asymptotically stable. \label{ii of thrm: relation}
\end{enumerate}
\end{theorem}
We outline the proof of the above theorems in Appendix \ref{appen: proof of theorems}. For detailed definitions of the terms used above, see \cite{2006-Chicone-ODE}. Roughly speaking, Theorem \ref{thrm: relation} states that under certain conditions (i.e., if the equilibrium points of systems \eqref{eq: Swing Equation Net Preserving} and \eqref{eq: Swing Equation Net Preserving singular perturbation} are hyperbolic for any small $\varepsilon$), then the stability properties of an equilibrium point of system \eqref{eq: Swing Equation Net Preserving} is the same as those of system \eqref{eq: Swing Equation Net Preserving singular perturbation}. Therefore, we can confidently use the results of Theorem \ref{thrm: main result}, as the stability certificate in this theorem will also guarantee the stability of the original structure-preserving system \eqref{eq: Swing Equation Net Preserving}.
%
\subsection{Fast and distributed scheme for stability assessment}
\label{subsec: Fast and distributed scheme}

The proposed control scheme is based on Theorem \ref{thrm: main result}. Specifically, condition \eqref{eq: condition for stability} offers a distributed control rule instructing how to change the operating point and parameters of the system in order to move towards stability. For our purposes, it is convenient to reorder the terms in \eqref{eq: condition for stability} and define the stability index
\begin{align} \label{eq: stability index}
            C_i = -Q_i - V_i^2B_{ii} - \frac{d_i^2}{2m_i}     && \forall i \in \mathcal{G}.
\end{align}
The proposed scheme works as follows: Using local measurements of reactive power $Q_i$ and voltage $V_i$, each generator computes the value of $C_i$ for itself. If each generator makes sure its $C_i$ is nonpositive, then the small-signal stability of the entire system is guaranteed.

Note that the proposed scheme is totally distributed and does not need any information from the neighboring generators. This property makes it suitable for fast small-signal stability assessment in large-scale power systems. We will show in Section \ref{subsec: Distributed stability framework} that $C_i$ can be used as an stability index, that is, as $C_i$ moves towards $-\infty$, the system roughly speaking becomes more stable (the real part of eigenvalues of the system moves towards $-\infty$).

A more conservative stability certificate will also be presented in the next section in Corollary \ref{coro: paradox}. According to this corollary, the small-signal stability can be certified based only on the local network topology information. This criterion is useful for topology design and planning problems, where system operators only have limited information about the operating point of the system.

\subsection{Remarks on Theorem \ref{thrm: main result}}

First and foremost, condition \eqref{eq: condition for stability} in Theorem \ref{thrm: main result} revolves only around the generator buses, confirming that small-signal stability is concerned with the rotor angle stability of the generators.

The variable $Q_i$ in \eqref{eq: condition for stability} is the net reactive power injected from bus $i$ into the network, that is, if the generator at bus $i$ is supplying reactive power, then $Q_i>0$. Otherwise, if it is consuming reactive power, then $Q_i<0$. Intuitively, when the generator at bus $i$ is a supplier of reactive power, the first term on the left-hand side of \eqref{eq: condition for stability} is negative, and this situation will help condition \eqref{eq: condition for stability} hold, thereby improving the stability of the system.

Recall that $Y_{ii} \angle \theta_{ii}=G_{ii}+jB_{ii} = \sum_{j=1}^n y_{ij}$, where $y_{ij}=g_{ij}+jb_{ij}$ is the admittance of line $(i,j)$, with $g_{ij}\ge0$ and $b_{ij}\le0$. Therefore, $B_{ii}\le0$, and the second term on the left-hand side of \eqref{eq: condition for stability} is always positive. Here, it is assumed that $y_{ii}$, i.e., the admittance-to-ground at bus $i$ is negligible. Otherwise, we may have $B_{ii}>0$, and the second term on the left-hand side of \eqref{eq: condition for stability} could be negative.

Condition \eqref{eq: condition for stability} enforces an upper bound which is proportional to the square of damping and inverse of inertia. This is consistent with the intuition that if we increase the damping, the stability margin of the system will increase. However, it is not intuitive (could be a paradox) that decreasing the inertia of a generator will increase the stability margin.

By adding more transmission lines to the system, $|B_{ii}|$ will increase, and this in turn could increase the left-hand side of \eqref{eq: condition for stability} and lead to instability. This can be called the Braess's Paradox \cite{2005Braess} in power system stability. The next corollary will further illustrate this stability paradox.
\begin{corollary}\label{coro: paradox}
Consider the singular perturbation model \eqref{eq: Swing Equation Net Preserving singular perturbation} with an equilibrium point $x^*$ that satisfies Assumption \ref{assump}. If the condition 
\begin{align} \label{eq: condition for stability without sin}
            \sum \limits_{j=1, j \ne i}^n { V_i V_j Y_{ij} } \le \frac{d_i^2}{2m_i},     && \forall i \in \mathcal{G}
\end{align}
is satisfied, then the equilibrium point is locally asymptotically stable.
\end{corollary}
This corollary directly follows from the proof of Theorem \ref{thrm: main result} provided in Appendix \ref{appen: proof of theorems}. Counterintuitively, according to \eqref{eq: condition for stability without sin}, adding more power lines can lead to violating the sufficient condition for stability and making the system unstable. This Braess's Paradox in power systems has been also acknowledged for example in \cite{2015-robustness-paradox-network, 2019-Balestra-Multistability, 2018-Tchuisseu-paradox} in different context and using different approaches.

\section{Numerical experiments}
\label{sec: Numerical experiments}
\begin{figure}[]
    \centering
	\begin{tikzpicture}
	\draw
    (0,0) node [sin v source] (v1){}   
    (v1.east)--++(0.25,0) coordinate(bus2) 
    (bus2)--++(0,0.4) node[right]{$2$}
    (bus2)--++(0,-0.4)
    (bus2)--++(0.6,0) coordinate(T2-L)  
    (T2-L)++(0.35,0) coordinate(T2-R)
    (T2-R)--++(0.6,0) coordinate(bus7)  
    (bus7)--++(0,0.4) node[right]{$7$}
    (bus7)--++(0,-0.2) coordinate(bus7-r)
    (bus7-r)--++(0,-0.2)
    (bus7)--++(1.5,0) coordinate(bus8)  
    (bus8)--++(0,0.4) node[right]{$8$}
    (bus8)--++(0,-0.2) coordinate(bus8-load)
    (bus8-load)--++(0.25,0) coordinate(bus8-load-right)
    (bus8-load)--++(0,-0.2)
    (bus8)--++(1.5,0) coordinate(bus9)  
    (bus9)--++(0,0.4) node[right]{$9$}
    (bus9)--++(0,-0.2) coordinate(bus9-l)
    (bus9-l)--++(0,-0.2)
    (bus9)--++(0.6,0) coordinate(T3-L)  
    (T3-L)++(0.35,0) coordinate(T3-R)
    (T3-R)--++(0.6,0) coordinate(bus3)  
    
    (bus3)--++(0,0.4) node[left]{$3$}
    (bus3)--++(0,-0.4)
    (bus3)--++(0.25,0) node[right,sin v source]{}
    (bus7-r)--++(0.25,0)--++(0,-1) coordinate(bus5)
    (bus5)--++(0.2,0) node[right]{$5$}
    (bus5)--++(-0.4,0) coordinate(bus5-load)
    (bus5-load)--++(-0.2,0)
    (bus9-l)--++(-0.25,0)--++(0,-1) coordinate(bus6)
    (bus6)--++(0.4,0) coordinate(bus6-load)
    (bus6-load)--++(0.2,0) 
    (bus6)--++(-0.2,0) node[left]{$6$}
    (bus5)--++(0,-0.25)--++(1.05,-0.6)--++(0,-0.25) coordinate(bus4-from5)
    (bus6)--++(0,-0.25)--++(-1.05,-0.6)--++(0,-0.25) coordinate(bus4-from6)
    (bus4-from5)--++(0.2,0)  coordinate(bus4)
    (bus4-from5)--++(-0.2,0)
    (bus4-from6)--++(0.2,0) node[right]{$4$}
    (bus4-from6)--++(-0.2,0)
    
    (bus4)--++(0,-0.6) coordinate(T1-N)  
    (T1-N)++(0,-0.35) coordinate(T1-S)
    (T1-S)--++(0,-0.6) coordinate(bus1)  
    
    (bus1)--++(0.4,0) 
    (bus1)--++(-0.4,0) node[left]{$1$}
    (bus1)--++(0,-0.25) node[below,sin v source]{}
    ;
    \draw[-Latex] (bus5-load) --++ (0,-0.6);
    \draw[-Latex] (bus6-load) --++ (0,-0.6);
    \draw[-Latex] (bus8-load-right) --++ (0,-0.6);
    
    \draw (T2-L) arc (-90:90:0.15);
    \draw (T2-L) arc (90:-90:0.15);
    \draw (T2-R) arc (90:270:0.15);
    \draw (T2-R) arc (270:90:0.15);
    \draw (T3-L) arc (-90:90:0.15);
    \draw (T3-L) arc (90:-90:0.15);
    \draw (T3-R) arc (90:270:0.15);
    \draw (T3-R) arc (270:90:0.15);
    \draw (T1-N) arc (-180:0:0.15);
    \draw (T1-N) arc (0:-180:0.15);
    \draw (T1-S) arc (0:180:0.15);
    \draw (T1-S) arc (180:0:0.15);
    \end{tikzpicture}
	\caption{Single line diagram of the WSCC system.}
	\label{fig: WSCC}      
\end{figure}
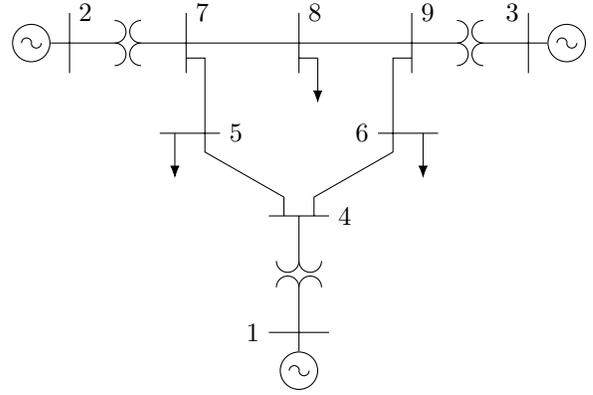

Consider the popular western system coordinating council (WSCC) $9$-bus $3$-generator system \cite{matpower}, depicted in Fig. \ref{fig: WSCC}. The base MVA is $100$, the system frequency is $60$ Hz, the network has nonzero transfer conductances, and the line complex powers are around hundreds of MVA each.

In Sections \ref{subsec: Time-domain comparison} and \ref{subsec: Modal comparison}, we verify Theorem \ref{thrm: relation} by showing that the singular perturbation model \eqref{eq: Swing Equation Net Preserving singular perturbation} can be used instead of the network-preserving model \eqref{eq: Swing Equation Net Preserving} for stability analysis. Then, in Section \ref{subsec: Distributed stability framework}, we show the application of Theorem \ref{thrm: main result} in fast and distributed stability assessment.

%
\subsection{Perturbed model approximation: Time-domain comparison}
\label{subsec: Time-domain comparison}
\begin{figure}[t]
	\centering
	\begin{subfigure}{0.49\textwidth}
		\centering
		\includegraphics[width=\linewidth, keepaspectratio=true]{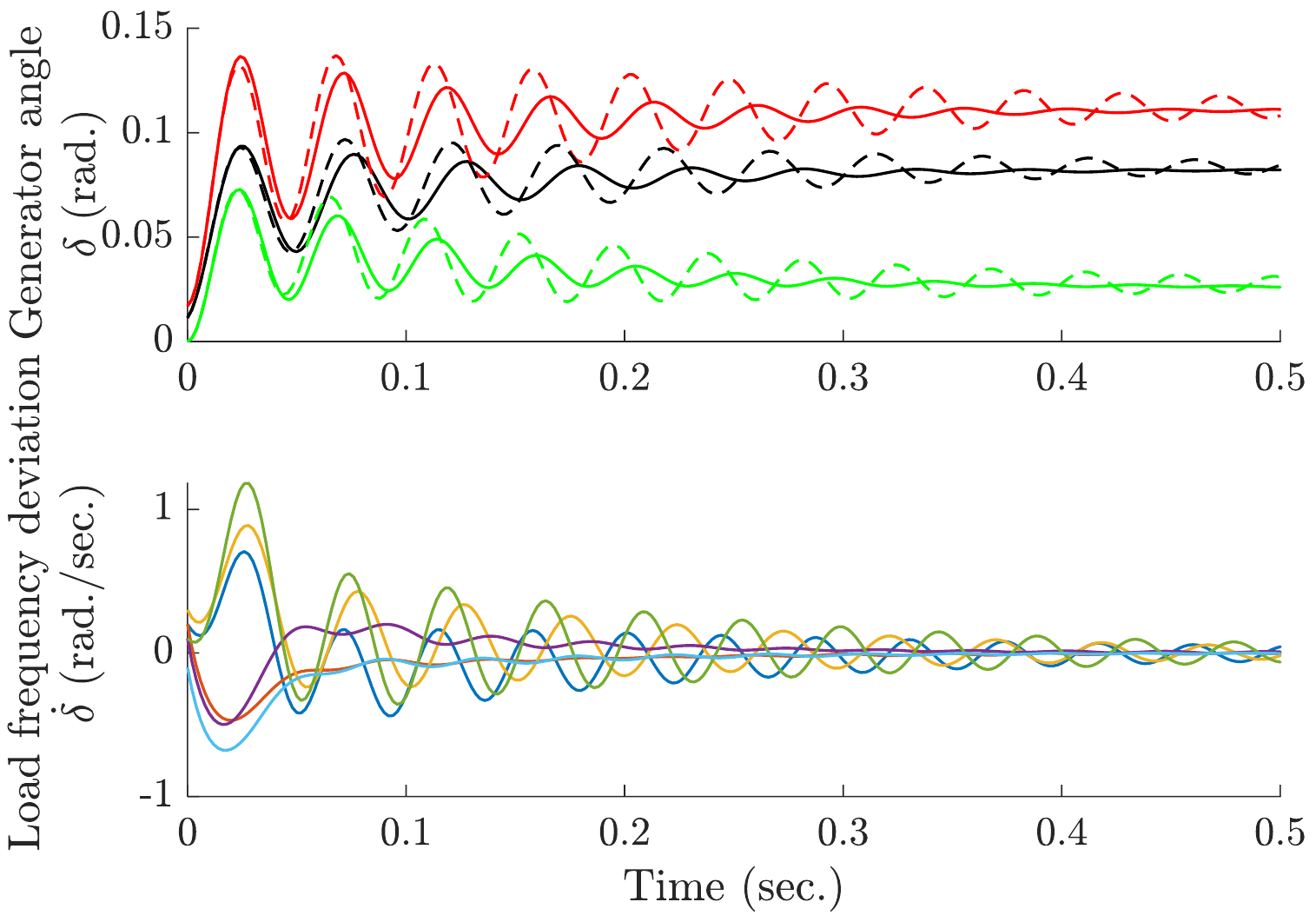}
		\caption{Using perturbation parameter $\varepsilon=10^{-2}$.}
	\end{subfigure}%
	
	\begin{subfigure}{0.49\textwidth}
		\centering
		\includegraphics[width=\linewidth, keepaspectratio=true]{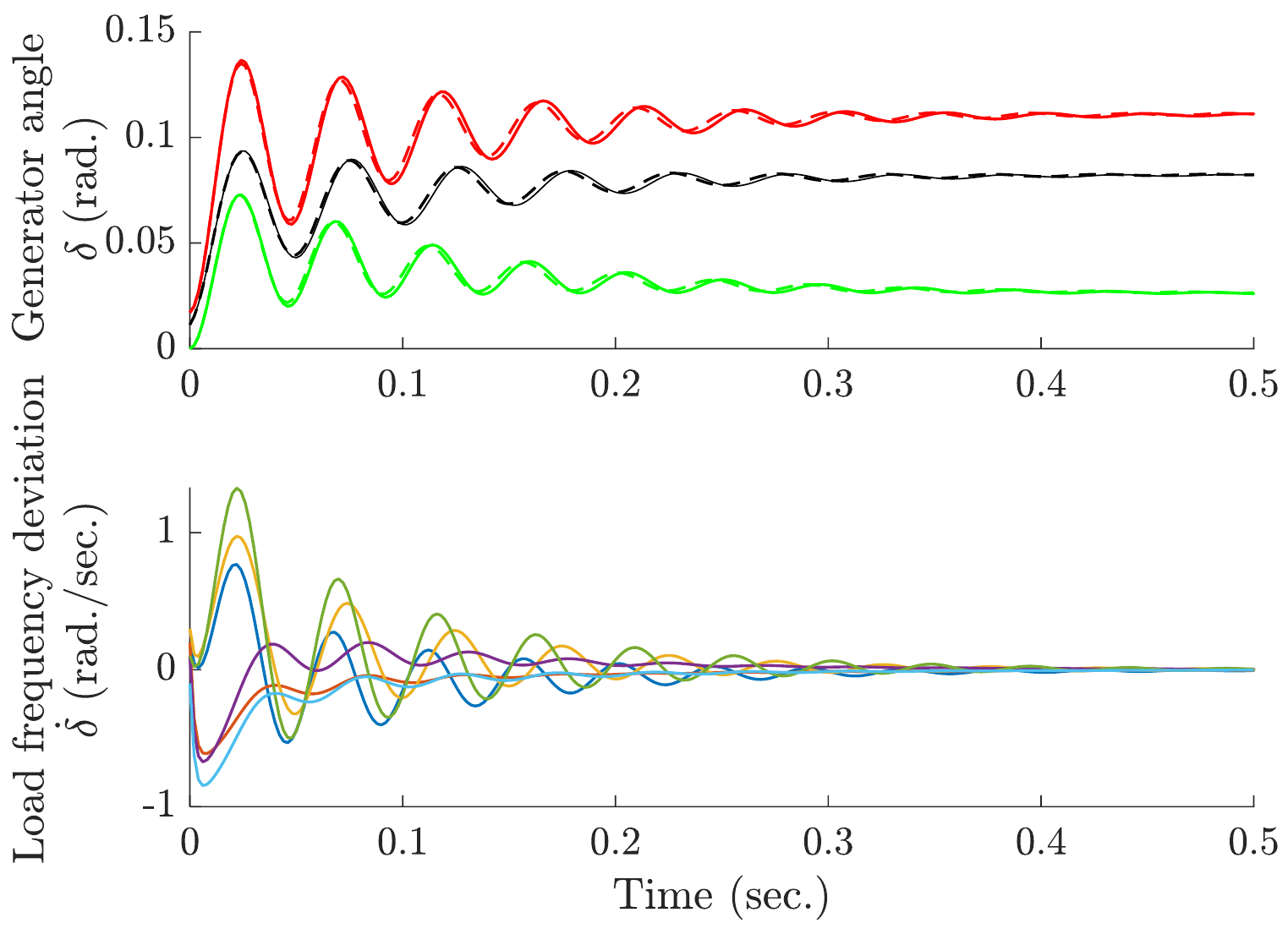}
		\caption{Using perturbation parameter $\varepsilon=2\times10^{-3}$.}
	\end{subfigure}
	\caption{Simulation results of the WSCC system: exact model (solid) and singular perturbation model (dashed) with two different perturbation parameters.}
	\label{fig: num_sim_WSCC}
\end{figure}

The singular perturbation model \eqref{eq: Swing Equation Net Preserving singular perturbation} can be viewed as an approximation of the network-preserving model \eqref{eq: Swing Equation Net Preserving}. In Fig. \ref{fig: num_sim_WSCC}, the upper figure in each subfigure (a) and (b) compares the voltage angle $\delta_i$ at generator buses of the exact (solid) structure-preserving model with those of the approximate (dashed) singular perturbation model. The trajectories of the two models clearly converge to the same stable equilibrium point, confirming Theorem \ref{thrm: relation}. Moreover, as we decrease the perturbation parameter $\varepsilon$ from $10^{-2}$ in subfigure (a) to $2\times10^{-3}$ in subfigure (b), the approximation error also decreases. Indeed, it can be rigorously proved that the aforementioned estimation is $\mathcal{O}(\varepsilon)$.

Fig. \ref{fig: num_sim_WSCC} also shows the frequency deviation $\dot \delta_i$ at load buses of the singular perturbation model. Recall that these $\dot{\delta}_i,\forall i \in \mathcal{L}$ were the state variables whose time derivative was multiplied by $\varepsilon$ in \eqref{eq: Swing Equation Net Preserving singular perturbation loads}. From \eqref{eq: Swing Equation Net Preserving singular perturbation loads}, the time derivative of $\dot{\delta}_i$ at load buses is $\ddot{\delta}_i  = (-{{d}}_i \dot{\delta}_i - P_{d_i} - P_{e_i})/{\varepsilon}$, which can be large when $\varepsilon$ is small. Accordingly, in Fig. \ref{fig: num_sim_WSCC} (b) with a smaller $\varepsilon$ compared to Fig. \ref{fig: num_sim_WSCC} (a), the dynamics of $\dot{\delta}_i$ at load buses converges more rapidly to zero.

%
%
\subsection{Perturbed model approximation: Modal analysis}
\label{subsec: Modal comparison}

\begin{figure}[t]
	\centering
	\begin{subfigure}{0.49\textwidth}
		\centering
		\includegraphics[width=\linewidth, keepaspectratio=true]{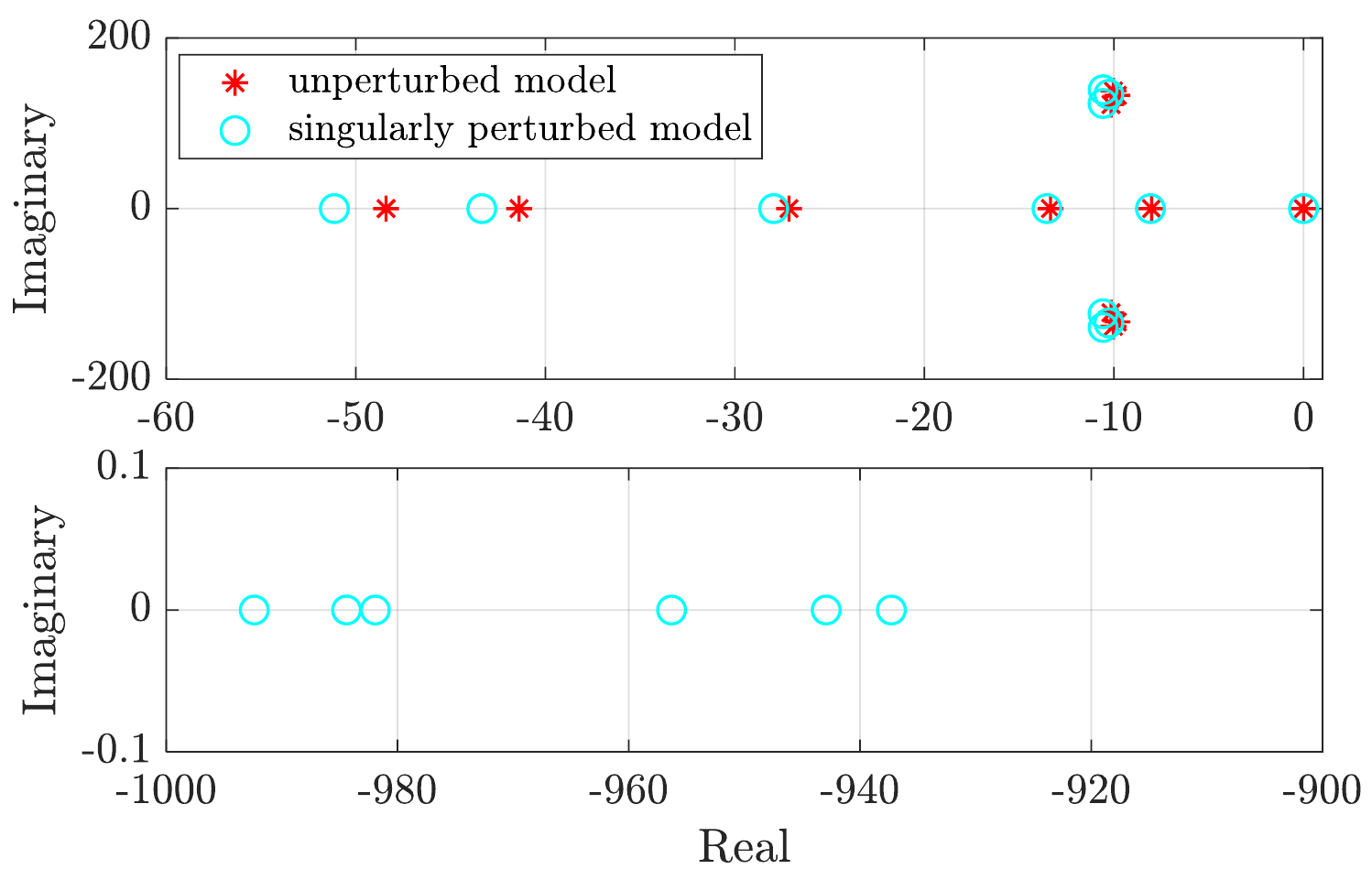}
		\caption{Using perturbation parameter $\varepsilon=10^{-3}$.}
	\end{subfigure}%
	
	\begin{subfigure}{0.49\textwidth}
		\centering
		\includegraphics[width=\linewidth, keepaspectratio=true]{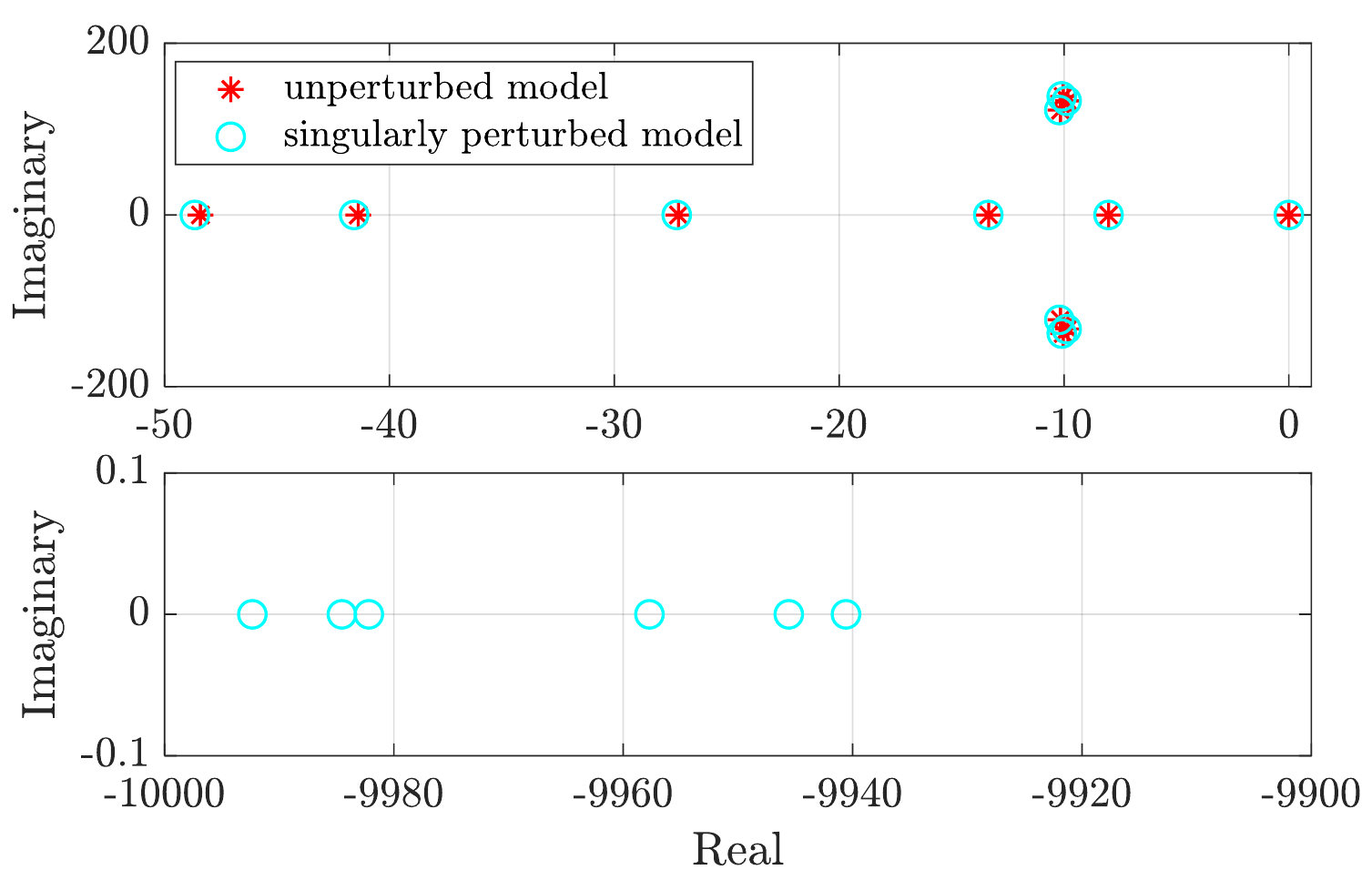}
		\caption{Using perturbation parameter $\varepsilon=10^{-4}$.}
	\end{subfigure}
	\caption{Modal analysis of the WSCC system: eigenvalues of the exact model (red asterisks) and singular perturbation model (cyan circles) with two different perturbation parameters.}
	\label{fig: modal_WSCC}
\end{figure}

Fig. \ref{fig: modal_WSCC} provides a comparison between the eigenvalues associated with the Jacobian matrix of models \eqref{eq: Swing Equation Net Preserving} and \eqref{eq: Swing Equation Net Preserving singular perturbation}. The two models have a set of eigenvalues which are close to each other. Additionally, note that the state space of the singular perturbation model has more dimensions (in this WSCC example, it has $6$ additional dimensions which is equal to the number of load buses). These additional eigenvalues are also shown separately in each subfigure.

Comparing Figs. \ref{fig: modal_WSCC}(a) and \ref{fig: modal_WSCC}(b), as the perturbation parameter gets smaller, the set of eigenvalues of model \eqref{eq: Swing Equation Net Preserving} approaches those of model \eqref{eq: Swing Equation Net Preserving singular perturbation}. Moreover, using a smaller perturbation parameter, the additional eigenvalues of the singular perturbation model move towards $-\infty$. Indeed, as $\varepsilon\to0$, the two systems will have a set of common eigenvalues, while the additional eigenvalues of the singular perturbation model will approach $-\infty$. Finally, observe that as $\varepsilon\to0$, the eigenvalues of the singular perturbation model do not approach the imaginary axis. According to Theorem \ref{thrm: relation}, the equilibrium points of the two models \eqref{eq: Swing Equation Net Preserving} and \eqref{eq: Swing Equation Net Preserving singular perturbation} have the same stability properties. This justifies the use of model \eqref{eq: Swing Equation Net Preserving singular perturbation} instead of model \eqref{eq: Swing Equation Net Preserving} for stability assessment.

%
%

\subsection{Fast and distributed stability assessment}
\label{subsec: Distributed stability framework}

\begin{figure}[]
    \centering
	\includegraphics[width=0.8\linewidth]{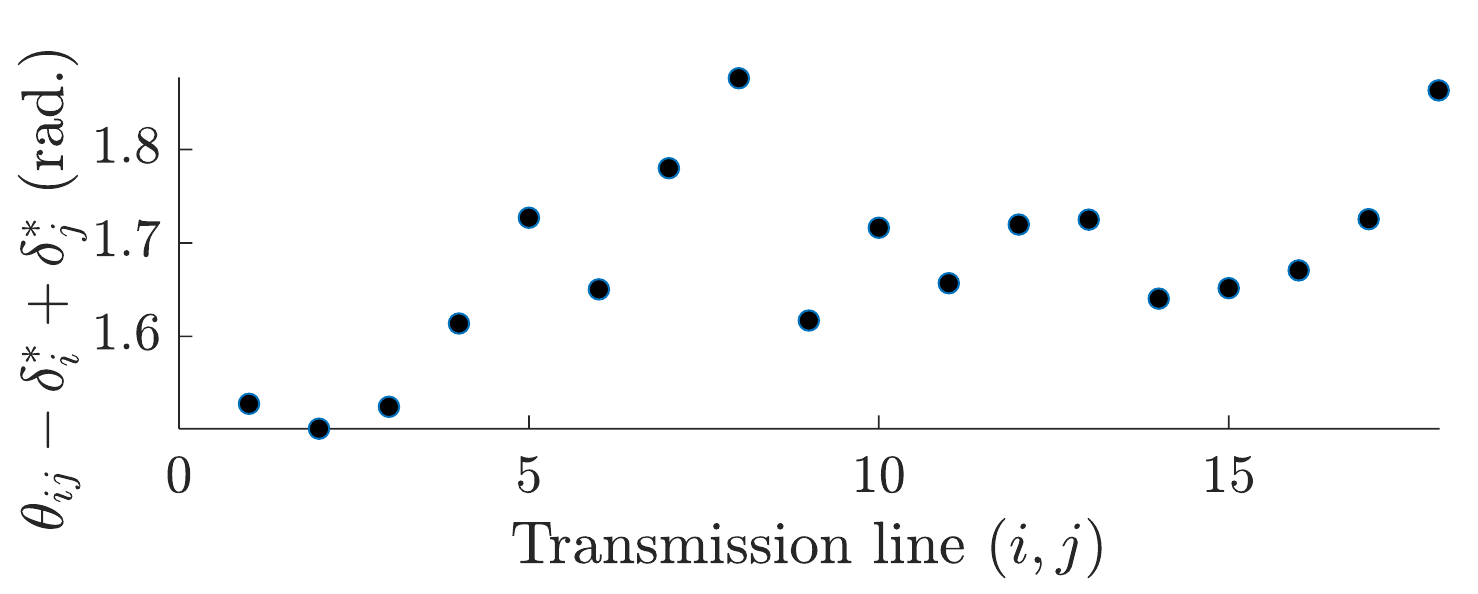}
	\caption{Verification of Assumption \ref{assump} in the WSCC system.}
	\label{fig: angle assump}      
\end{figure}

As mentioned previously, Assumption \ref{assump} is reasonable and holds in practice. Fig. \ref{fig: angle assump} confirms this issue for the WSCC system. As can be seen, the angles $(\theta _{ij} - \delta _i^* + \delta _j^*) $ for all transmission lines are perfectly located within the interval $0$ to $\pi$ rad. Moreover, Theorem \ref{thrm: relation} has been verified in Sections \ref{subsec: Time-domain comparison} and \ref{subsec: Modal comparison}, thereby justifying the use of Theorem \ref{thrm: main result} and the singular perturbation model for stability assessment. In this section, we test the efficacy of the scheme proposed in Section \ref{subsec: Fast and distributed scheme}.

Recall when the stability index $C_i$ defined in \eqref{eq: stability index} is negative for all generators, then by Theorem \ref{thrm: main result}, the equilibrium point of the system is asymptotically stable. Note that the converse may not be true, i.e., $C_i$ could be positive while the system is stable. However, even in such cases, $C_i$ can be viewed as an index, showing the degree of stability.

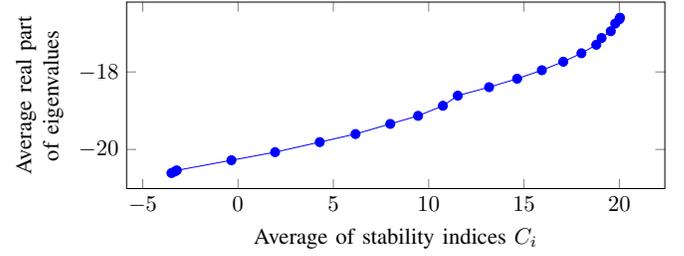
\begin{figure}[]
\centering
\begin{tikzpicture}[scale=0.85]
	\begin{axis}[
		width=10cm,
	    height=4.5cm,
		xlabel=Average of stability indices $C_i$,
		ylabel style={align=center},
		ylabel={Average real part \\ {of eigenvalues} }]
	\addplot[color=blue,mark=*] coordinates {
		(20.034,-16.6)
        (19.999,-16.643)
        (19.783, -16.753)
        (19.550, -16.950)
        (19.066,-17.125)
        (18.789, -17.300)
        (18.011,-17.518)
        (17.060, -17.737)
        (15.936, -17.956)
        (14.639, -18.175)
        (13.169, -18.393)
        (11.526, -18.612)
        (10.745, -18.873)
        (9.443, -19.133)
        (7.985, -19.341)
        (6.162, -19.602)
        (4.287,-19.810)
        (1.943,-20.070)
        (-0.348,-20.279)
        (-3.212,-20.539) 
        (-3.324, -20.571)
        (-3.492, -20.608)
	};
	\end{axis}
\end{tikzpicture}
	\caption{Variation of degree of stability due to variation of stability index \eqref{eq: stability index}.}
    \label{fig: ave real vs index}
\end{figure}

Consider the WSCC system under different operating points as well as different system parameters (generators' inertia and damping). As the operating points or system parameters vary, the eigenvalues of the system may also move to either right half-plane (less stable) or left half-plane (more stable). Now, the stability index \eqref{eq: stability index} helps us understand how the eigenvalues move if we vary operating points or system parameters. Fig. \ref{fig: ave real vs index} shows the variation in the real parts of eigenvalues of model \eqref{eq: Swing Equation Net Preserving singular perturbation} as a function of changes in the stability index \eqref{eq: stability index}. In this figure, under all operating conditions and system parameters, the equilibrium point is asymptotically stable. However, as the average of stability indices moves towards negative value (i.e., the violation of condition \eqref{eq: condition for stability} decreases and at some point the condition holds), the average real part of eigenvalues move towards $-\infty$, making the operating point more stable.
\section{Conclusions}
\label{sec: Conclusions}
We showed under reasonable assumptions, the small-signal stability of the classical structure-preserving model is equivalent to its singular perturbation counterpart. Based on this equivalence, we developed a novel stability certificate for the structure-preserving model. The certificate can be computed in a fully distributed fashion, using only local information, and can be used for real-time monitoring. The certificate suggests that the eigenvalues of the system will move towards the left half-plane by increasing generators’ damping and decreasing generators’ inertia. It also reveals a paradox that adding more transmission lines can lead to the violation of the stability certificate and making the system unstable. The stability certificate could be incorporated as a constraint into various problems such as the optimal power flow problem in order to guarantee and improve the stability of solutions. Our results could also be extended towards tighter and nonlocal stability certificates.
%
%

\appendices
\section{Proofs of Theorems \ref{thrm: main result} and \ref{thrm: relation}}
\label{appen: proof of theorems}
The structure-preserving model \eqref{eq: Swing Equation Net Preserving} can be written as the following system of first-order differential equations:
\begin{subequations} \label{eq: SE NP reduced}
	\begin{align}
	\dot{\delta}_i & = \omega_i && \forall i \in \mathcal{G},  \\
	{{d}}_i \dot{\delta}_i & =  -P_{d_i} - P_{e_i} && \forall i \in \mathcal{L},  \\
	m_i \dot{\omega}_i &=    -d_i{\omega}_i  +  P_{m_i}  -  P_{e_i}  && \forall i \in \mathcal{G},
	\end{align}
\end{subequations}
where $\omega_i$ is the deviation of angular frequency from its nominal value. Similarly, the singular perturbation model \eqref{eq: Swing Equation Net Preserving singular perturbation} can be written as
\begin{subequations} \label{eq: Swing Equation Net Preserving first order singular perturbation}
	\begin{align}
	\dot{\delta}_i & = \omega_i && \forall i \in \mathcal{G},  \\
	\dot{\delta}_i & = \omega_i && \forall i \in \mathcal{L},  \\
	 m_i \dot{\omega}_i  &= - d_i {\omega}_i  + P_{m_i}  - P_{e_i}  && \forall i \in \mathcal{G},  \\
	 \varepsilon {\dot{\omega}_i} &= - {{d}}_i \omega_i - P_{d_i} - P_{e_i}  && \forall i \in \mathcal{L}.
	\end{align}
\end{subequations}
In the sequel, we use ${{m}}_i$ as an alias for $\varepsilon, \forall i \in \mathcal{L}$ in order to represent its physical interpretation. In other words, ${{m}}_i=\varepsilon, \forall i \in \mathcal{L}$. Let us define 
$D=\mathbf{diag}(d_1,\cdots,d_{n_0}, {d}_{n_0+1},\cdots,{d}_n),$ and $M=\mathbf{diag}(m_1,\cdots,m_{n_0}, {m}_{n_0+1},\cdots,{m}_n)$. The Jacobian of \eqref{eq: Swing Equation Net Preserving first order singular perturbation} is 
\begin{align}\label{eq:J}
J = \begin{bmatrix}
0 & I \\
-M^{-1} L     &     -M^{-1} D\\
\end{bmatrix},
\end{align}
where $L$ is the Jacobian of the flow function \eqref{eq: flow function}. Now, we are ready to present an outline of the proof of Theorems \ref{thrm: main result} and \ref{thrm: relation}.
\subsection{Outline of the proof of Theorem \ref{thrm: main result}}
The proof consists of the following steps \cite{2020-fast-certificate}:
\begin{enumerate}
    \item Under Assumption \ref{assump}, the spectrum of matrix $L$ is in the right half-plane. Moreover, $L$ has a simple zero eigenvalue.
    \item $\lambda\in\mathbb{C}$ is an eigenvalue of $J$ if and only if the quadratic matrix pencil $L+\lambda D + \lambda^2 M$ is singular.
    \item Using Gershgorin circle theorem and the above steps, we can show that all nonzero real eigenvalues of $J$ are negative.
    \item Let $\lambda\in\mathbb{C}$ be an eigenvalue of $J$. According to step $2$ of this proof, there exists a nonzero vector $v$ such that $(L + \lambda D + \lambda^2 M ) v = 0$. Normalize vector $v$ such that $\max |v_i| = 1$, and let $k=\argmax |v_i|$. 
    
    \item Spelling out the $k$th row of $(L + \lambda D + \lambda^2 M ) v = 0$, we have 
    \begin{align} \label{eq: kth row}
        [L]_{kk}v_k + \lambda D_k v_k + \lambda^2 M_kv_k   = -\sum_{i\ne k} [L]_{ki}v_i
    \end{align}
    
    \item Suppose condition \eqref{eq: condition for stability} holds. Assume for the sake of contradiction that $\lambda$ has a positive real part. Use \eqref{eq: kth row} and lead it to the contradiction that a positive number is less than or equal to a nonpositive number.
\end{enumerate}
See \cite{2020-fast-certificate} for the details.

\subsection{Outline of the proof of Theorem \ref{thrm: relation}}

We follow the Tikhonov’s theorem \cite{1999-khalil-kokotovic-singular-perturbation}:
\begin{enumerate}
    \item Define a reference bus, and write the referenced swing equation model. This will put us in a convenient position to apply Tikhonov’s theorem.
    \item The boundary layer model associated with the singular perturbation model \eqref{eq: Swing Equation Net Preserving first order singular perturbation} can be simplified to the linear differential equation
    \begin{align} \label{eq: SE boundary layer model thrm 2}
        \frac{dy_i}{d\tau} = - {{d}}_i y_i && \forall i \in \mathcal{L}, 
    \end{align}
    where $\tau$ is the the new time variable defined as $\tau = \frac{t-t_0}{\varepsilon}$,
and $y_i$ is defined as
\begin{align}
    y_i = \omega_i + \frac{P_{d_i}}{{{d}}_i} + \frac{1}{{{d}}_i} P_{e_i}, && \forall i\in \mathcal{L}.
\end{align}
Therefore, the origin is a globally exponentially stable equilibrium point of this boundary layer model as ${{d}}_i>0, \forall i \in \mathcal{L}$. Now, Statement (\ref{i of thrm: relation}) of Theorem \ref{thrm: relation} follows from \cite[Section 7, Corollary 2.3]{1999-khalil-kokotovic-singular-perturbation}.
\item Let $K$ be the Jacobian of the first order system \eqref{eq: SE NP reduced}. Show that $K$ is a Schur complement of $J$. Then, show if $K$ has $r$ eigenvalues with negative real part, then there exits a sufficiently small $\varepsilon$ such that $J$ also has $r$ eigenvalues with negative real part. 

\item To prove Statement (\ref{ii of thrm: relation}) of Theorem \ref{thrm: relation}, assume for the sake of contradiction that $x^*$ is not an asymptotically stable equilibrium point of \eqref{eq: SE NP reduced}. Since $x^*$ is hyperbolic, there must exist an eigenvalue in the right half-plane. Using step $3$ of this proof, we reach the contradiction that system \eqref{eq: Swing Equation Net Preserving first order singular perturbation} is not asymptotically stable. 

\bibliographystyle{IEEEtran}
\bibliography{main}

\end{enumerate}






\end{document}

%% file: main-packages.tex
\usepackage{amsmath}  

\usepackage{arydshln}
\usepackage{dsfont}

\usepackage[]{algorithm2e}
\usepackage{makeidx}
\usepackage{amsthm,amssymb}
\usepackage{bbold}
\usepackage{latexsym,remreset}
\usepackage{graphicx}
\usepackage{multirow}
\usepackage{bbm}
\usepackage{amsfonts}
\usepackage{pgfplots}
\pgfplotsset{compat=newest}
\usepackage{mathtools}
\usepackage{textcomp}
\usepackage{scrextend}
\usepackage{commath}
\usepackage[stable]{footmisc}

\usepackage{graphicx}
\usepackage{url}
\usepackage{enumerate}

\usepackage{subcaption}

\usepackage{tikz}
\usepackage{cite}

\makeatletter
\def\url@leostyle{%
	\@ifundefined{selectfont}{\def\UrlFont{\sf}}{\def\UrlFont{\small\ttfamily}}}
\makeatother
\newtheorem{theorem}{Theorem}

\newtheorem{assumption}[]{Assumption}

\newtheorem{corollary}[]{Corollary}

\DeclareMathOperator*{\argmax}{argmax}

\newcommand*{\rom}[1]{\expandafter\@slowromancap\romannumeral #1@}

\usepackage{color}
\definecolor{mygreen}{RGB}{10, 128, 128}
\definecolor{myspia}{RGB}{0, 100, 0}
\usepackage{hyperref}
\hypersetup{
	colorlinks   = true, 
	urlcolor     = black, 
	linkcolor    = myspia, 
	citecolor   = myspia 
}
\usepackage{pgfplotstable}
\usepackage{pgfplots}
\pgfplotsset{compat=1.8}
\usepgfplotslibrary{statistics}
\usepackage{tikz}
\usetikzlibrary{calc}
\usetikzlibrary{arrows}
\usetikzlibrary{arrows,decorations.markings}
\usetikzlibrary{arrows.meta}
\usepackage[american]{circuitikz}

\tikzset{sin v source/.style={
  circle,
  draw,
  append after command={
    \pgfextra{
    \draw
      ($(\tikzlastnode.center)!0.5!(\tikzlastnode.west)$)
       arc[start angle=180,end angle=0,radius=0.425ex] 
      (\tikzlastnode.center)
       arc[start angle=180,end angle=360,radius=0.425ex]
      ($(\tikzlastnode.center)!0.5!(\tikzlastnode.east)$) 
    ;
    }
  },
  scale=1.5,
 }
}